\documentclass{article}
\usepackage{acra}
\usepackage{graphicx}
\usepackage{amsmath,amssymb,amsfonts}
\usepackage{algorithmic}
\usepackage{textcomp}
\usepackage{xcolor}
\usepackage{booktabs}
\usepackage{balance}
\usepackage{fancyhdr}  

\title{Self context-aware emotion perception on human-robot interaction}

\author{%
  \fontsize{12pt}{14pt}\selectfont
  Zihan Lin\textsuperscript{1}, Francisco Cruz\textsuperscript{1,2}, Eduardo Benitez Sandoval\textsuperscript{3}\\
  \fontsize{12pt}{14pt}\selectfont
  \textsuperscript{1}School of Computer Science and Engineering, University of New South Wales, Sydney, Australia\\
  \textsuperscript{2}Escuela de Ingenier\'ia, Universidad Central de Chile, Santiago, Chile\\
   \textsuperscript{3}School of Art and Design, Creative Robotics Lab, University of New South Wales, Sydney, Australia\\
  Emails: \texttt{zihan.lin3@student.unsw.edu.au, f.cruz@unsw.edu.au, e.sandoval@unsw.edu.au}
}

\usepackage{fancyhdr}  

\fancypagestyle{firstpage}{  
    \fancyhf{}  
    \fancyhead[L]{\small \centering \noindent Cite as: Zihan Lin, Francisco Cruz, and Eduardo Benitez Sandoval. Self context-aware emotion perception on human-robot interaction. In Proceedings of the Australasian Conference on Robotics and Automation (ACRA). 2023.}  
}

\begin{document}
\thispagestyle{firstpage}

\maketitle

\begin{abstract}
Emotion recognition plays a crucial role in various domains of human-robot interaction. In long-term interactions with humans, robots need to respond continuously and accurately, however, the mainstream emotion recognition methods mostly focus on short-term emotion recognition, disregarding the context in which emotions are perceived. Humans consider that contextual information and different contexts can lead to completely different emotional expressions. In this paper, we introduce self context-aware model (SCAM) that employs a two-dimensional emotion coordinate system for anchoring and re-labeling distinct emotions. Simultaneously, it incorporates its distinctive information retention structure and contextual loss. This approach has yielded significant improvements across audio, video, and multimodal. In the auditory modality, there has been a notable enhancement in accuracy, rising from 63.10\% to 72.46\%. Similarly, the visual modality has demonstrated improved accuracy, increasing from 77.03\% to 80.82\%. In the multimodal, accuracy has experienced an elevation from 77.48\% to 78.93\%. In the future, we will validate the reliability and usability of SCAM on robots through psychology experiments.  
\end{abstract}

\section{Introduction}
Human-robot interaction has become increasingly important due to the widespread use of robots in various applications such as manufacturing, healthcare, and personal assistance \cite{kyrarini-etal:robots-healthcare}. Human-robot interaction focuses on how humans and robots can safely and effectively collaborate, requiring natural and intuitive communication between them. To achieve better communication, it is crucial for robots to understand human emotions; otherwise, they may respond incorrectly, leading humans to reject interacting with the robots \cite{tsiourti-etal:multimodal-emotion-integration}. However, emotion recognition is a complex process, involving various perceptual dimensions and temporal aspects. Current emotion recognition models primarily focus on multimodal perception but often overlook contextual information \cite{poria-etal:emotion-recognition}. In real-life conversational contexts, emotional fluctuations in individuals often display a sense of continuity. This implies that, under typical circumstances, emotions do not undergo abrupt and dramatic shifts within a brief timeframe, such as sudden transitions from intense anger to extreme happiness. Consequently, when emotions cannot be ascertained, humans frequently depend on contextual information to make judgments \cite{sacharin-etal:changing-emotion-expressions}. Practically, there is often a requirement to rely exclusively on preceding contextual information.

Therefore, this paper introduces self context-aware model (SCAM), enabling a robot to perform emotion recognition on the user while simultaneously considering the user's preceding emotional context and integrating it with the robot's recognition results from the preceding context. This approach allows for a more comprehensive and accurate assessment of the user's emotion state during human-robot interactions.

The contributions of this work are summarized as follows:
\begin{itemize}

\item We utilize the relationship between valence, arousal, and emotion to enable the model to learn basic emotions from non-basic ones.
\item We introduce a novel contextual loss, incorporating the model's predictions of context emotion, valence, and arousal, allowing the model to more effectively capture emotional change trends.
\item We model the information transfer within the context, preserving valuable features from the preceding context for integration and judgment when making predictions for the subsequent context.
\item We conduct experiments on the IEMOCAP dataset, encompassing speech modality, visual modality, and multimodal scenarios. The experimental results demonstrate that our approach achieves significant improvements across all modalities, some of which have reached the state of the art.

\end{itemize}

\section{Related work} 

The application of emotion recognition spans a wide range of domains, including its deployment in various human-computer interaction scenarios and chatbot systems designed to generate emotionally rich dialogues \cite{zhou-etal:emotional-chatting-machine}. Nevertheless, this field is fraught with numerous challenges. For instance, individuals experiencing mental distress may be reluctant to unveil their vulnerabilities, often concealing their true emotional states \cite{maithri-etal:automated-emotion-recognition}. Consequently, researchers have delved extensively into the realm of emotion recognition, exploring various modalities such as visual, auditory, physical, and even the incorporation of EEG and skin conductance signals \cite{li-etal:eeg-emotion-recognition}. Typically, amalgamating information from diverse modalities yields enhanced accuracy in emotion recognition \cite{wu-etal:eeg-functional-connectivity}. Notably, the fusion of speech and text modalities achieved a remarkable accuracy rate of 80.51\% (four categories) on the IEMOCAP dataset \cite{atmaja-etal:bimodal-speech-emotion-recognition}.

However, some studies \cite{tsiourti-etal:multimodal-emotion-integration} underscore the intricacy of emotion recognition when information conflicts arise between modalities. Furthermore, some researchers suggested that better results can be achieved by capturing contextual information within conversations.\cite{priyasad-etal:attention-driven-fusion} employed graph neural networks to model inter-dialogue relationships, achieving commendable performance. They introduced an iterative emotion interaction network that employs iteratively predicted emotion labels to explicitly model emotion interactions, culminating in an accuracy of 64.37\% (seven categories) on the IEMOCAP dataset. Compared to providing a dialogue-based approach, we take into consideration that during human-robot interaction, the robot may struggle to provide sufficient feedback. Therefore, we propose a method to utilize self context.

Another pivotal dimension of research in emotion recognition is the exploration of dimensional emotion models. In contrast to discrete emotion models,  valence-arousal model \cite{russell1980circumplex} offers a better understanding of the intricate relationships among different emotional states. Research demonstrated that combining valence, arousal, dominance, and the polarity of emotions within a multi-view training framework can yield superior results \cite{tompkins-etal:multi-view-learning}. However, the current use of dimensional models has not effectively leveraged the continuity of dimensional models in emotional expression, failing to fully exploit their advantages. In comparison to discrete emotion models, dimensional models can better observe the trend of emotional changes. In our approach, we make the first attempt to utilize this aspect and have achieved excellent results.

\section{Methods}
\begin{figure}[htbp]
    \centering
    \includegraphics[width=0.4\textwidth]{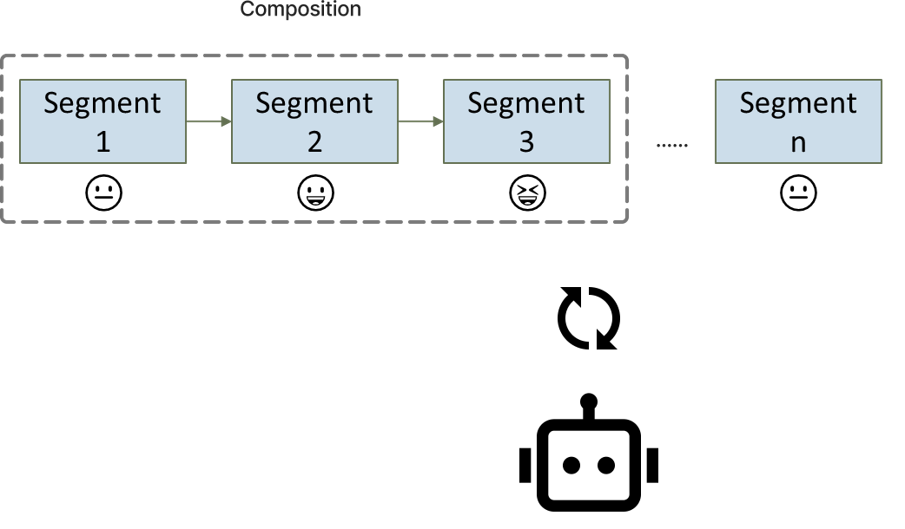}
    \caption{Context interaction in HRI} 
    \label{HRI}
\end{figure}

During the process of human-robot interaction, long-lasting sessions can be divided into multiple segments, and they may vary with their own context and the robot's responses. The emotions in each segment are relatively independent, with a correlation observed between adjacent segments due to the continuous nature of human emotions. Therefore, we group some adjacent segments into a composition (Figure \ref{HRI}). By using SCAM to capture this correlation, we achieve better emotion perception results.

SCAM consists of two main components: a multi-task network for each segment and a self context-aware structure for composition. The multi-task network relabels emotions and then utilizes ResNet101 and Bi-LSTM to recognize emotion, valence, and arousal within a short time. The self context-aware structure incorporates contextual information propagation and context loss, combining the context information and predictions from preceding segments with the current input to predict the current emotion, valence, and arousal. In the following sections, we will provide further details on these components.
\subsection{Dataset}
Considering there is currently no publicly available and suitable human-robot interaction dataset,  we conduct model training and validation on IEMOCAP dataset \cite{busso-etal:iemocap}, which comprises 10,039 instances performed by 10 actors. These actors are paired, and each pair engages in multiple scripted and spontaneous emotional dialogues, which are appropriate for the simulation of human-robot interaction.  Throughout these dialogues, they portray 10 predefined emotional states: angry, sad, happy, neutral, disgust, surprise, fear, excited, other, and unmarked. Each emotional state includes multiple sentences and encompasses various modalities, including audio, video, text, and more.

\begin{table}[h]
\centering
\renewcommand{\arraystretch}{1.2}
\begin{tabular}{ccc}
\hline
Emotion & Number of Samples & Rate (\%) \\
\hline
Anger & 1103 & 24.57 \\
Happy & 595 & 13.25 \\
Neutral & 1708 & 38.04 \\
Sadness & 1084 & 24.14 \\
\hline
\textbf{Total} & \textbf{4490} & \textbf{100.00} \\
\hline
\end{tabular}
\caption{IEMOCAP: Four Emotions}
\label{table:IEMOCAP-Four-Emotions}
\end{table}

\begin{itemize}
    \item Frame Segmentation: Divide the audio signal into small segments, 
    \item Windowing: Apply a window function to each frame to reduce the impact of spectral leakage.
    \item Fourier Transform: Apply the Discrete Fourier Transform (DFT) to each window, transforming the time-domain signal into a frequency-domain signal.
    \item Magnitude and Squaring: Compute the amplitude spectrum for each frequency component, often by taking the magnitude of the complex values.
    \item Visualization: Display the obtained spectral information as an image, with the horizontal axis representing time, the vertical axis representing frequency, and color or brightness representing amplitude.
\end{itemize}

Four emotions (angry, happy, neutral, and sad) and two modalities (auditory and visual) in IEMOCAP are used to verify our approach in order to compare with other methods, as shown in Table \ref{table:IEMOCAP-Four-Emotions}. Emotions are annotated using two mainstream approaches. One approach considers emotions as fixed labels, where each emotion is treated as a category such as angry, happy, sad, or neutral. Each sample is assigned only one emotion label, representing a specific emotion category. The other approach \cite{russell1980circumplex} involves emotion dimensions, where each sample has two labels annotated with valence and arousal. Valence represents one dimension of emotion (e.g., emotional intensity), and arousal represents another dimension (e.g., positivity and negativity). This approach better captures complex emotional changes, as emotions can be regarded as points in an emotion space rather than single categories.

\begin{figure}[htbp]
    \centering
    \includegraphics[width=0.35\textwidth]{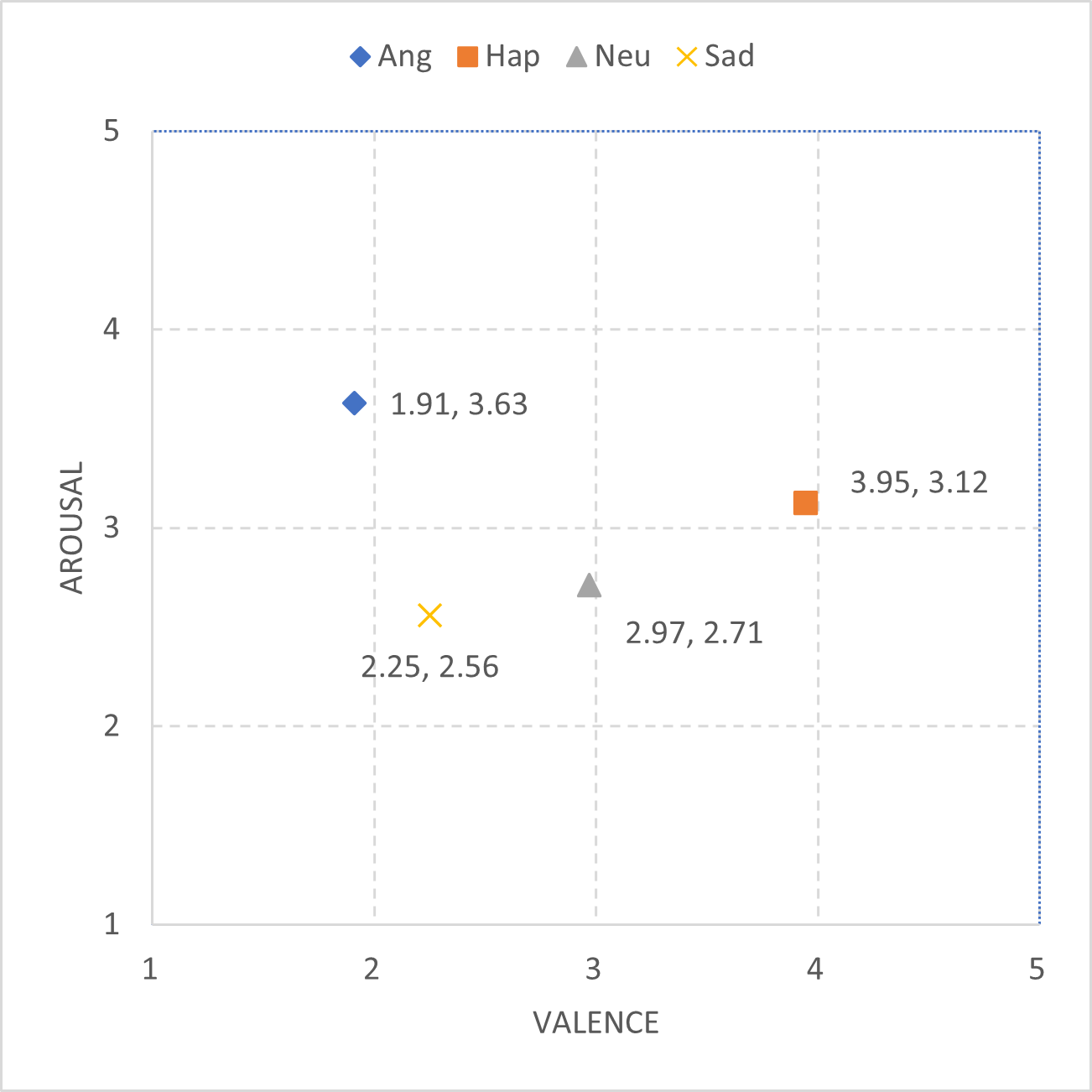}
    \caption{IEMOCAP emotions on Valence-Arousal axis} 
    \label{img5}
\end{figure}

We calculate the mean of valence and arousal as coordinates for each emotion in the emotion space, as shown in Figure \ref{img5}. Within the same composition, emotions tend to undergo a transition and be closer to the emotion of the current segment (the last segment in the composition). Capturing such contextual emotion changes can improve the continuity and accuracy of emotion perception.

Frame Segmentation: Divide the audio signal into small segments, 
Windowing: Apply a window function to each frame to reduce the impact of spectral leakage.
Fourier Transform: Apply the Discrete Fourier Transform (DFT) to each window, transforming the time-domain signal into a frequency-domain signal.
Magnitude and Squaring: Compute the amplitude spectrum for each frequency component, often by taking the magnitude of the complex values.
Visualization: Display the obtained spectral information as an image, with the horizontal axis representing time, the vertical axis representing frequency, and color or brightness representing amplitude.

\subsection{Multi-task network}

\subsubsection{Preprocessing}
\begin{figure}[htbp]
    \centering
    \includegraphics[width=0.3\textwidth]{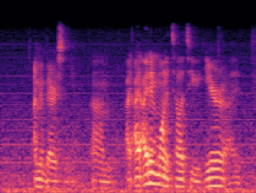}
    \caption{Log-Mel spectrogram of one segment} 
    \label{spectrogram}
\end{figure}

For different modalities of input, we employ distinct preprocessing methods. For the auditory modality, we convert speech into log-Mel spectrograms as input. Compared to one-dimensional audio signals, log-Mel spectrograms better represent crucial features in speech signals, such as formants and harmonic structures, enhancing the accuracy of emotion perception. The log-Mel spectrograms are generated with a sampling rate of 22050Hz and utilize 256 Mel filters. An example of a log-Mel spectrogram generated from a single segment is shown in Figure \ref{spectrogram}.

 \begin{figure}[htbp]
    \centering
    \includegraphics[width=0.5\textwidth]{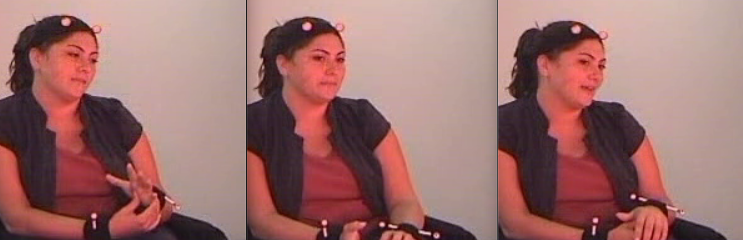}
    \caption{Cropped frames of one segment} 
    \label{faces}
\end{figure}

For the visual modality, considering the short duration of each segment and the limited facial expression changes, three frames are recorded. We extract the start, intermediate, and end frames from the corresponding video segments and crop the facial regions (Figure \ref{faces}). The intermediate frame is calculated using the start and end times of each segment.

\subsubsection{Segment structure}
\begin{figure}[htbp]
    \centering
    \includegraphics[width=0.5\textwidth]{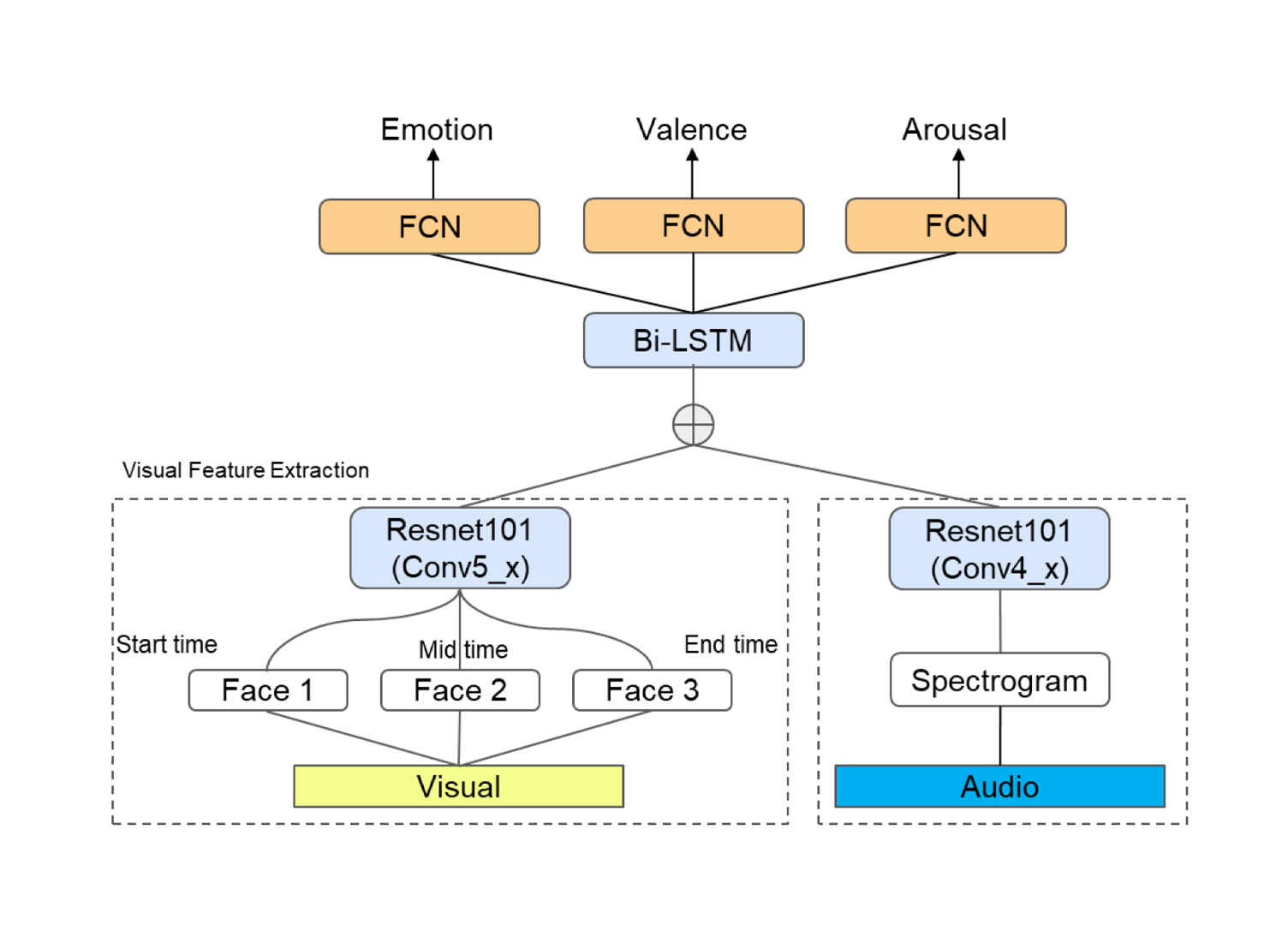}
    \caption{Segment structure (multimodal)} 
    \label{multitask}
\end{figure}

ResNet101 is utilized for feature extraction. After the ResNet101, high-level features are fed into a Bi-LSTM to capture temporal information. For the auditory modality, high-level features are extracted 
after Conv4\_x. For the visual modality, the three frames are separately fed into ResNet101, and after Conv5\_x, they are concatenated and passed to the Bi-LSTM. In the case of multi-modal data, feature alignment and concatenation occur before input to the Bi-LSTM. The Bi-LSTM's output is then split into three fully connected neural network (FCN), responsible for emotion classification, valence regression, and arousal regression, as shown in Figure \ref{multitask}.

\subsubsection{Segment Relabeling}

In the context of our contextual emotion model, each composition comprises multiple segments, and our objective is to predict the emotion of the last segment in each composition (the emotion at the current time). Given that emotions tend to exhibit relatively small variations within a composition, we select the emotion of the last segment as the emotion label for the entire composition, while valence and arousal remain unchanged, as shown in Table \ref{table:time-segments}. This approach offers two advantages. Firstly, it effectively leverages data from emotion labels other than the four basic emotions, enriching the dataset. Secondly, one segment's emotion may lead to different current emotions in different compositions. In the following section, we will explain how we utilize this feature.

\begin{table}[h]
\centering
\begin{tabular}{ccccccc}
\toprule
Segment & Emotion & Valence & Arousal & Relabel \\
\midrule
11 & Angry & 1.5 & 4 & Angry \\
12 & Frustrated & 1.5 & 4 &  Angry\\
13 & Angry & 1.5 & 4.5 &  Angry\\
\bottomrule
\end{tabular}
\caption{Relabel of composition}
\label{table:time-segments}
\end{table}

\subsubsection{Loss in segment}
Due to the relabeling of emotions for segments within each composition, in most cases, relabeled emotions closely match or are similar to the original emotions, owing to the continuity of emotions. However, there are instances where inconsistencies in emotions arise. As mentioned before, we compute the average valence and arousal for different emotions, serving as reference points on the two-dimensional emotion coordinate system. Simultaneously, we retain the original valence and arousal labels for each segment. Therefore, we utilize the Euclidean distance to measure the distance between the relabeled emotions and the original emotions and scale the emotion loss accordingly, as indicated by the following formula:

\[
\mathcal{L}_{emo} = -R \cdot \frac{1}{N} \sum_{i=1}^{N} \sum_{j=1}^{4} y_{ij} \log(p(y_{ij})),
\]which \(N\) represents the number of samples, \(y_{ij}\) denotes the actual emotion labels, and \(p(y_{ij})\) represents the predicted probabilities of emotions by the model.

\(R\) is defined as:
\[
R = \frac{1}{\sqrt{(x_{\text{emo}}-x_{\text{seg}})^2+(y_{\text{emo}}-y_{\text{seg}})^2}}.
\]\(x_{emo}\) and \(y_{emo}\) represent the valence and arousal of the relabel emotion (e.g., \(x_{emo}\) and \(y_{emo}\) of anger), and \(x_{seg}\) and \(y_{seg}\) represent the label of valence and arousal of the segment.

For valence and arousal regression use mean squared error, the formulas are as follows,

\[\mathcal{L}_{val} = \frac{1}{N} \sum_{i=1}^{N} (x_{seg} - \hat{x}_{seg})^2,\]

\[\mathcal{L}_{aro} = \frac{1}{N} \sum_{i=1}^{N} (y_{seg} - \hat{y}_{seg})^2,\]which \(\hat{x}_{seg}\) and \(\hat{y}_{seg}\) represent the prediction of valence and arousal seperately.

\subsection{Self context-aware structure}

\begin{figure}[htbp]
    \centering
    \includegraphics[width=0.5\textwidth]{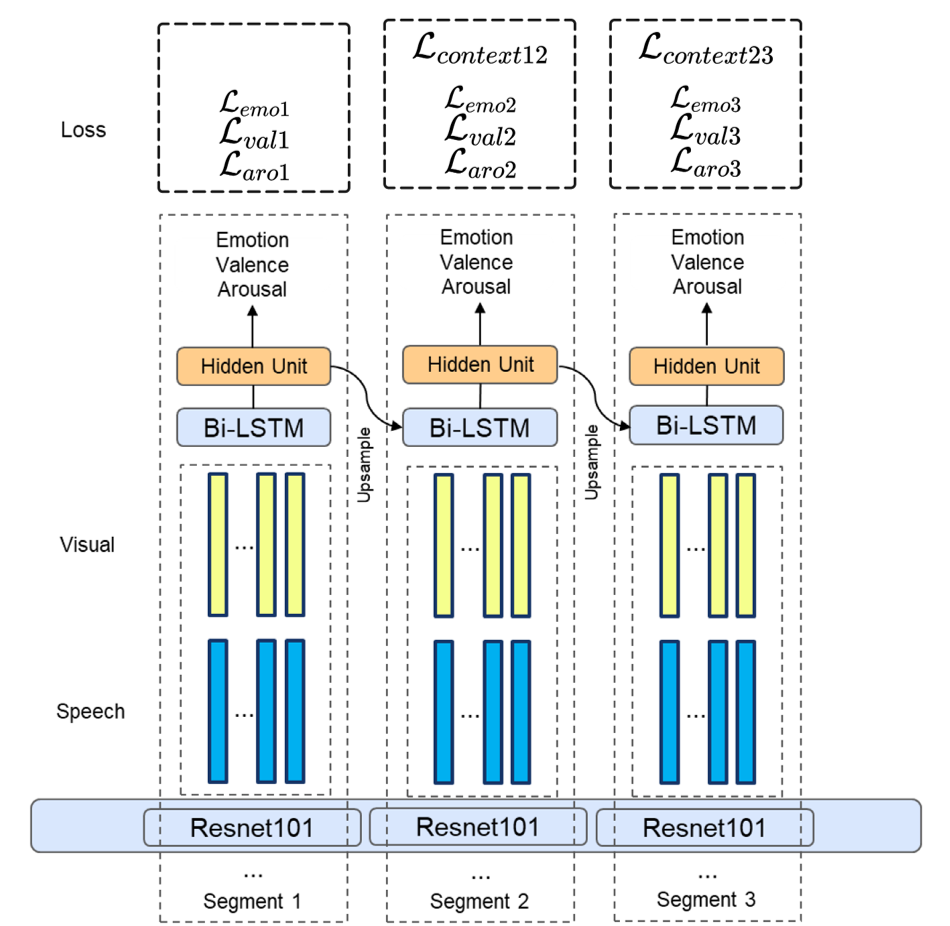}
    \caption{Context-aware structure} 
    \label{context_model}
\end{figure}

Self context-aware structure is the core optimization component of our model, primarily consisting of two parts: contextual information propagation and context loss. Through these components, we achieve the perception of current emotion with the aid of contextual information. The procedure is shown in Figure \ref{context_model}.

\subsubsection{Contextual Information Propagation}
Assume that the features processed by ResNet101 are denoted as $\{x_1^{(1)}, x_2^{(1)}, \ldots\}$, where $x_i^{(1)}$ represents the features of the $i$-th part in the first segment, and $\{x_1^{(2)}, x_2^{(2)}, \ldots\}$ represents the features of the second segment, and so on.

The output of the Bi-LSTM can be represented as $\{h_1^{(1)}, h_2^{(1)}, \ldots\}$, where $h_i^{(1)}$ denotes the output of the Bi-LSTM at the $i$-th time step in the first segment, and $\{h_1^{(2)}, h_2^{(2)}, \ldots\}$ represents the output of the Bi-LSTM in the second segment, and so forth. Since the Bi-LSTM is bidirectional, each $h_i^{(j)}$ contains both forward and backward propagation information and is typically represented as $h_i^{(j)} = [\overrightarrow{h_i^{(j)}}, \overleftarrow{h_i^{(j)}}]$, where $\overrightarrow{h_i^{(j)}}$ represents the forward propagation output, and $\overleftarrow{h_i^{(j)}}$ represents the backward propagation output.

Next, for each segment, we extend the output of the last time step to have the same dimensions as the input features for the LSTM time steps. This extension is represented as:

$$h_{\text{ext}}^{(j)} = U \cdot [\overrightarrow{h_{-1}^{(j)}}, \overleftarrow{h_{-1}^{(j)}}],$$
which $h_{\text{ext}}^{(j)}$ represents the extended output, and $U$ denotes an upsampling layer. $\overrightarrow{h_{-1}^{(j)}}$ represents the output of the last time step in the forward propagation for the $j$-th segment, and $\overleftarrow{h_{-1}^{(j)}}$ represents the output of the last time step in the backward propagation for the $j$-th segment.

Finally, we concatenate the extended output with all other features in the next segment, which is represented as:

$$\text{X}^{(j+1)} = [h_{\text{ext}}^{(j)}, x^{(j+1)}_1, x^{(j+1)}_2, \ldots],$$
which $\text{X}^{(j+1)}$ represents the input for the Bi-LSTM in the next segment, $h_{\text{ext}}^{(j)}$ is the extended output from the current segment, and $x^{(j+1)}_1, x^{(j+1)}_2, \ldots$ are all the input features in the next segment.

In this way, the model can utilize high-dimensional emotional information from the previous segment to better understand the emotions in the current segment.

\subsubsection{Context Loss}
The context loss is employed to capture emotional variations between adjacent segments. We represent the valence and arousal of each segment within a composition on a two-dimensional emotion coordinate system, as illustrated in Figure \ref{img5}. The vectors formed between consecutive segments depict the trends in emotional changes. We measure the distance between the predicted vectors and actual vectors using cosine similarity, thus forming the context loss, as shown below:

\[ \mathcal{L}_{context} = \frac{{\mathbf{v}^{ij}_{pre} \cdot \mathbf{v}^{ij}_{label}}}{{\| \mathbf{v}^{ij}_{pre} \| \cdot \| \mathbf{v}^{ij}_{label} \|}}, \text{ where } j - i = 1 .\]
\(i\) represents the \(i\)th segment, \(j\) represents the \(j\)th segment, and \(j - i = 1\) denotes adjacent segments.

The following notations are used:
\((x^{i}_{pre}, y^{i}_{pre})\) denotes the predicted coordinates of the valence and arousal for the \(i\)th segment. \((x^{j}_{pre}, y^{j}_{pre})\) denotes the predicted coordinates of the valence and arousal for the \(j\)th segment.
\((x^{i}_{label}, y^{i}_{label})\) denotes the true coordinates of the valence and arousal for the \(i\)th segment.
\((x^{j}_{label}, y^{j}_{label})\) denotes the true coordinates of the valence and arousal for the \(j\)th segment.

Since \(j - i = 1\), we can form two vectors    :
\(\mathbf{v}^{ij}_{pre}= (x^{j}_{pre} - x^{i}_{pre},\  y^{j}_{pre} - y^{i}_{pre})\), representing the predicted emotion change from \(i\) to \(j\).
 \(\mathbf{v}^{ij}_{label} = (x^{j}_{label} - x^{i}_{label},\  y^{j}_{label} - y^{i}_{label})\), representing the labeled emotion change from \(i\) to \(j\).
 \begin{figure}[htbp]
    \centering
    \includegraphics[width=0.35\textwidth]{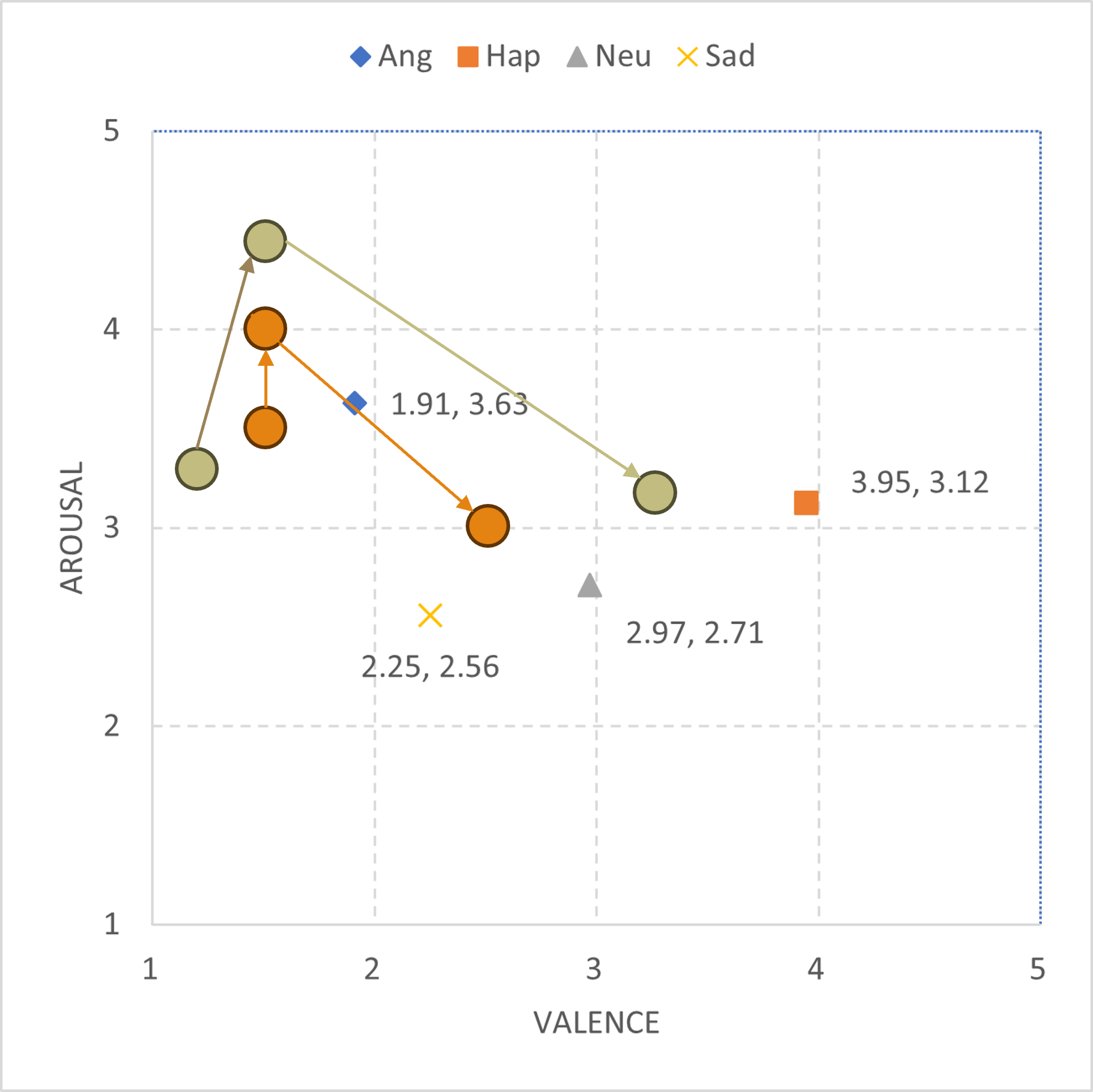}
    \caption{Context loss in valence-arousal axis} 
    \label{context_loss}
\end{figure}
As shown in Figure \ref{context_loss}, in which orange points represent label valence and arousal and the grey points represent predict valence and arousal, SCAM can learn the trend through context loss. If the cosine similarity is close to 1, it indicates that \(\mathbf{v}^{ij}_{pre}\) and \(\mathbf{v}^{ij}_{label}\) have similar directions, which means that the model's predictions are in line with the true labels. If the cosine similarity is close to -1, it indicates that \(\mathbf{v}^{ij}_{pre}\) and \(\mathbf{v}^{ij}_{label}\) have opposite directions, which means that there is a large deviation between the predictions and the true labels, indicating that the model does not capture the trend in the context effectively.

\section{Experiments}
\begin{figure}[htbp]
    \centering
    \includegraphics[width=0.4\textwidth]{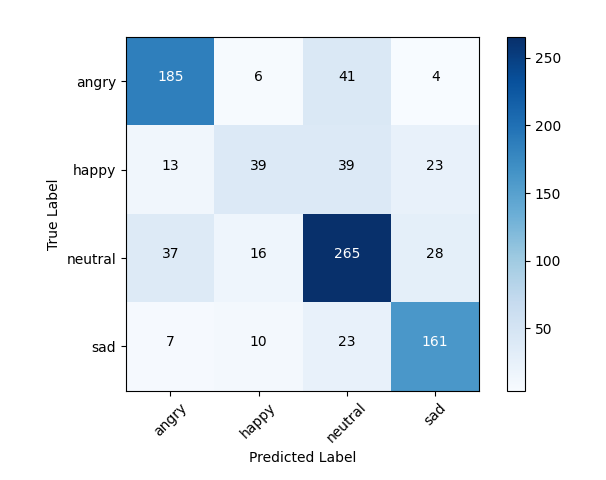}
    \caption{Confusion matrix of A-SCAM} 
    \label{speech_context}
\end{figure}
\subsection{Results}
Due to the random selection of the test dataset (10\% of IEMOCAP of four emotions), some segments have discontinuous context. In such cases, we replicate the current segment as a substitute for the missing context. We compare the results of the auditory modality (A-SCAM), visual modality (V-SCAM), and multimodal (M-SCAM) with the following baselines, as shown in Table \ref{table:model-comparison}.

\begin{figure}[htbp]
    \centering
    \includegraphics[width=0.4\textwidth]{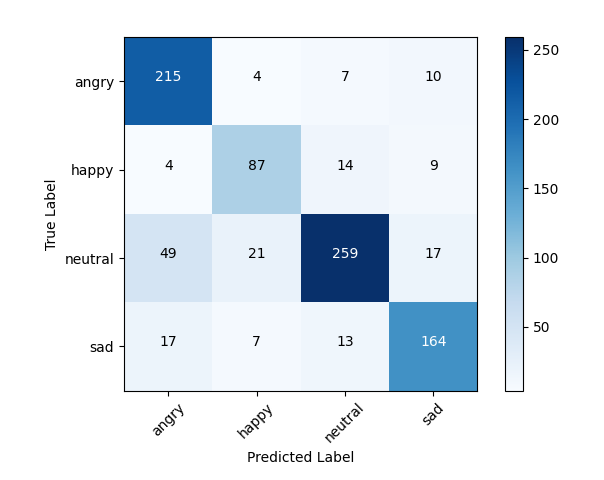}
    \caption{Confusion matrix of V-SCAM} 
    \label{visual_context}
\end{figure}

\begin{figure}[htbp]
    \centering
    \includegraphics[width=0.4\textwidth]{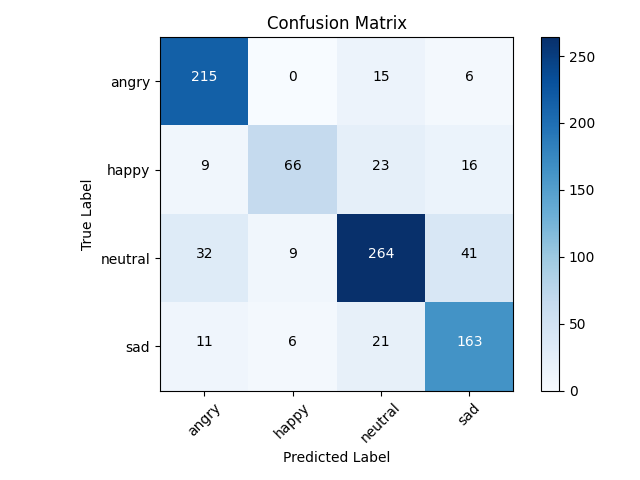}
    \caption{Confusion matrix of M-SCAM} 
    \label{multimodal_context}
    \end{figure}

In the auditory modality, SCAM achieves an accuracy of 72.46\%, in the visual modality, it reaches the highest accuracy of 80.82\%, and in the multimodal, it achieves an accuracy of 78.93\%. Though the multimodal performance is slightly inferior, SCAM achieves results superior to the baseline in both the auditory and visual modalities. The corresponding confusion matrices for different modalities are shown in Figure \ref{speech_context}, Figure \ref{visual_context}, and Figure \ref{multimodal_context}. In general, the visual modality performs better overall than the auditory modality, and the accuracy in recognizing the happy emotion is significantly higher in the visual modality.

\begin{table*}[h]
\centering
\resizebox{\textwidth}{!}{%
\begin{tabular}{cccccc}
\toprule
Model & Train/Predict & Modality & Emotion (UA) & Valence & Arousal \\
\midrule
\multicolumn{6}{c}{Prior Work} \\
\midrule
Audio-CNN-xvector\cite{peng-etal:efficient-speech-emotion-recognition} & Emo & A & 68.40 & - & - \\
SVM\cite{tompkins-etal:multi-view-learning} & V/A, Emo & A & 68.23 & - & - \\
w2v2-b\cite{tompkins-etal:multi-view-learning} & V/A/D, Emo & A & 61.20 & 47.80 & 60.50 \\
MMAN\cite{pan-etal:multi-modal-attention} & Emo & A+V+T & 73.94 & - & - \\
AV-ITN\cite{fu-etal:av-itn} & Emo & A+V & 81.66 & - & - \\
\midrule
\multicolumn{6}{c}{Proposed Work} \\
\midrule
\textbf{A-SCAM} & V/A, Emo & A & 72.46 & 53.18 & 67.78 \\
\textbf{V-SCAM} & V/A, Emo & V & \textbf{80.82} & \textbf{60.09} & 59.09 \\
\textbf{M-SCAM} & V/A, Emo & A+V & 78.93 & 59.87 & \textbf{68.78} \\
\bottomrule
\end{tabular}%
}
\caption{Performance comparison of prior work and SCAM}
\label{table:model-comparison}
\end{table*}

\subsection{Analysis}
\subsubsection{Efficiency of context loss}

We also evaluate the results when the loss is minimized on the test set, as shown in Table \ref{table:lowest-test-loss}.
\begin{table}[h]
\centering
\setlength{\tabcolsep}{5pt}
\renewcommand{\arraystretch}{1.2}
\begin{tabular}{cccc}
\toprule
Model & Emotion (UA) & Valence & Arousal \\
\midrule
A-SCAM & 60.98 & 49.50 & 64.33 \\
V-SCAM & 78.48 & 61.20 & 58.19 \\
M-SCAM & 76.48 & 57.64 & 69.01 \\
\bottomrule
\end{tabular}
\caption{Performance at lowest test loss}
\label{table:lowest-test-loss}
\end{table}

\begin{figure}[htbp]
    \centering
    \includegraphics[width=0.45\textwidth]{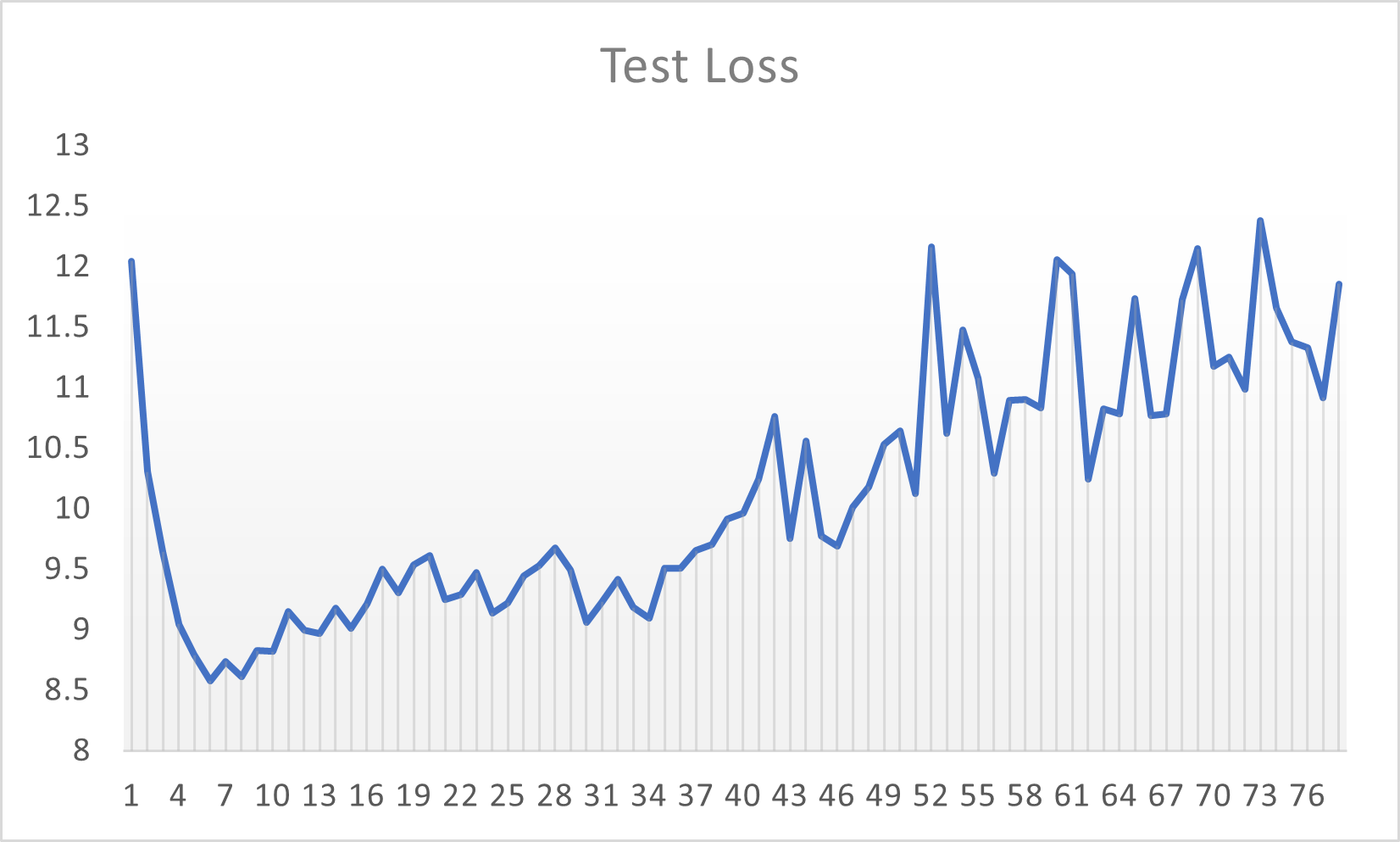}
    \caption{Test loss of auditory modality} 
    \label{test_loss}
\end{figure}

\begin{figure}[htbp]
    \centering
    \includegraphics[width=0.45\textwidth]{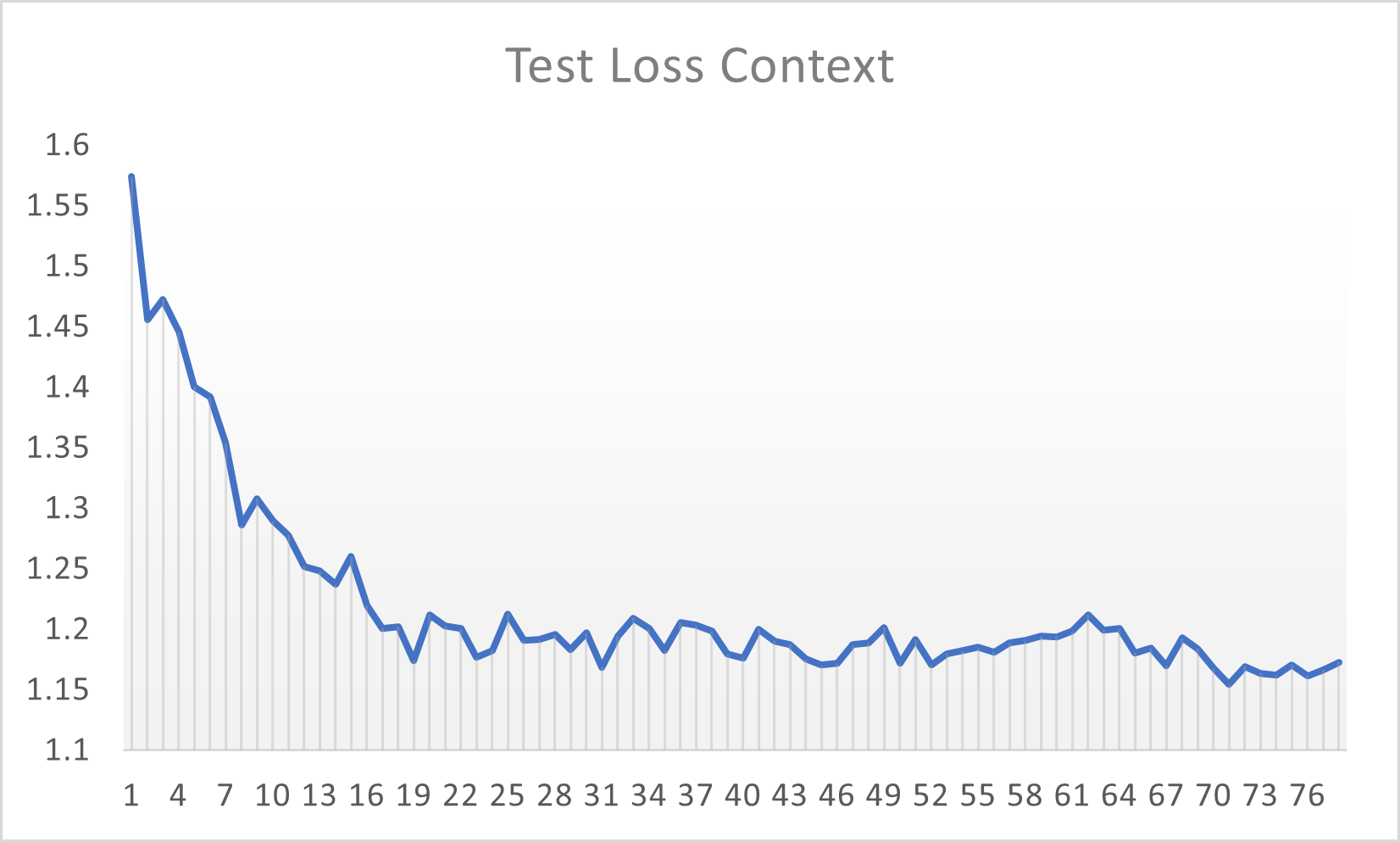}
    \caption{Test loss context of auditory modality} 
    \label{test_loss_context}
\end{figure}

\begin{figure}[htbp]
    \centering
    \includegraphics[width=0.45\textwidth]{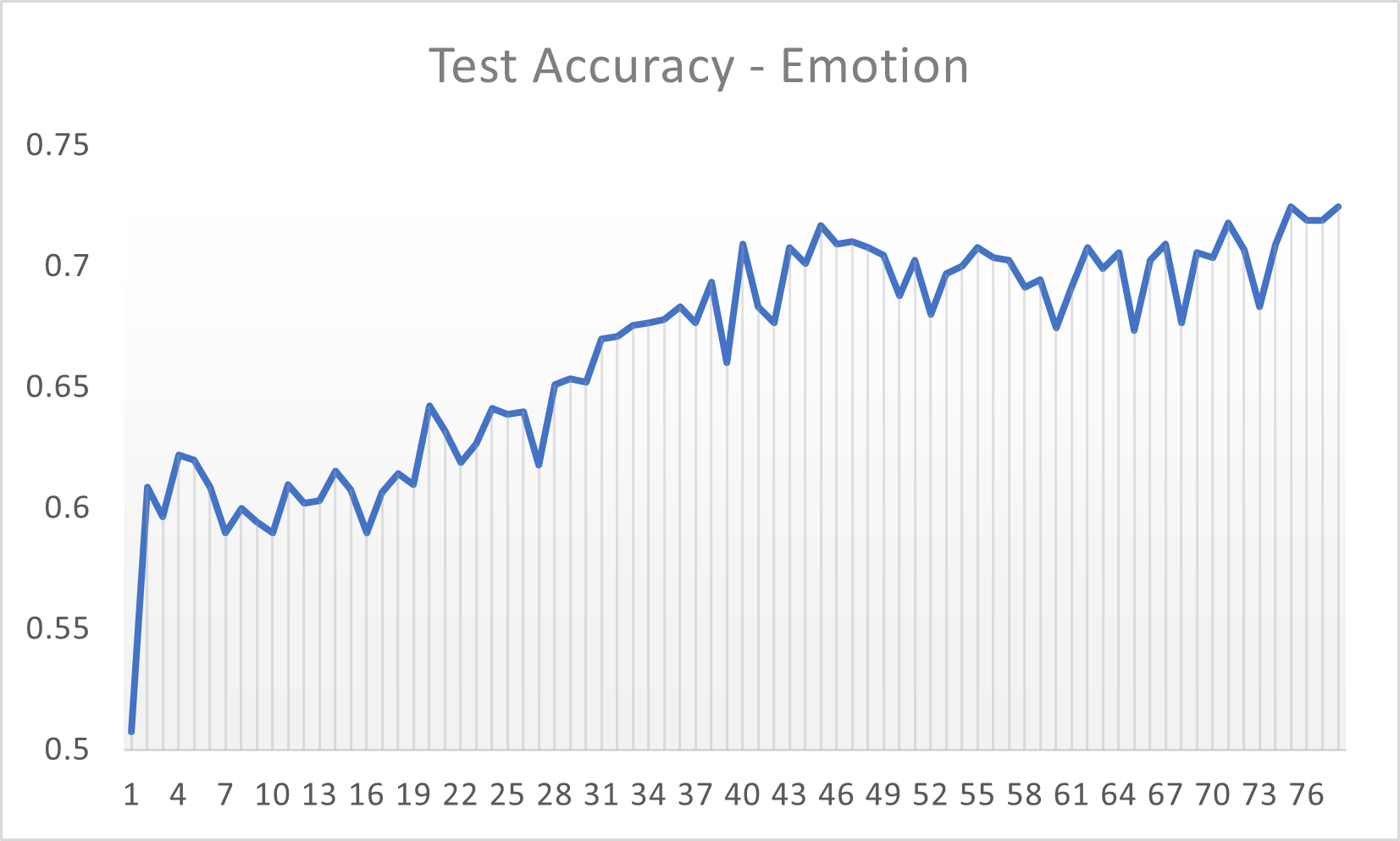}
    \caption{Test accuracy of emotion of auditory modality} 
    \label{test_acc_emotion}
\end{figure}

In our experiments, a difference in the trends of accuracy and loss is observed. The lowest loss occurrs significantly earlier than the highest accuracy. In the case of a single segment, this difference is primarily influenced by the multitask loss. Although valence and arousal are simultaneously predicted, the primary task remains emotion classification. Therefore, loss and accuracy may not be entirely correlated. For SCAM, the loss composition becomes more complex. It includes context loss due to the relabeling process and the necessity to predict the emotion of preceding segments in order to predict the current emotion. This context loss may not necessarily reflect the current emotion and may even conflict with the current emotion, valence, and arousal. Taking the auditory modality as an example, even though the total loss (Figure \ref{test_loss}) on the test set fluctuates and even increases, the context loss (Figure \ref{test_loss_context}) consistently decreases. This indicates that the model is effectively learning the contextual relationships of emotions, resulting in improved emotion classification results, as shown in Figure \ref{test_acc_emotion}


\subsubsection{Ablation Study}

In the case of using only a single segment, where emotional loss is not scaled and there is no self-context-aware structure, we compare the auditory modality (A-SEG), visual modality (V-SEG), and multimodal approach (M-SEG) with SCAM. The results are presented in Table \ref{table: ablation study}.

\begin{table}[h]
\centering
\setlength{\tabcolsep}{5pt}
\renewcommand{\arraystretch}{1.2}
\begin{tabular}{cccccc}
\toprule
Model & Emotion (UA) & Valence & Arousal \\
\midrule
A-SEG & 63.10 & 52.29 & 66.22 \\
A-SCAM & 72.46 & 53.18 & 67.78 \\
\midrule
V-SEG & 77.03 & 59.2 & 59.31 \\
V-SCAM & 80.82 & 60.09 & 59.09 \\
\midrule
M-SEG & 77.48 & 56.63 & 66.89 \\
M-SCAM & 78.93 & 59.87 & 68.78 \\
\bottomrule
\end{tabular}
\caption{Ablation experiments of SCAM}
\label{table: ablation study}
\end{table}

\begin{figure}[htbp]
    \centering
    \includegraphics[width=0.5\textwidth]{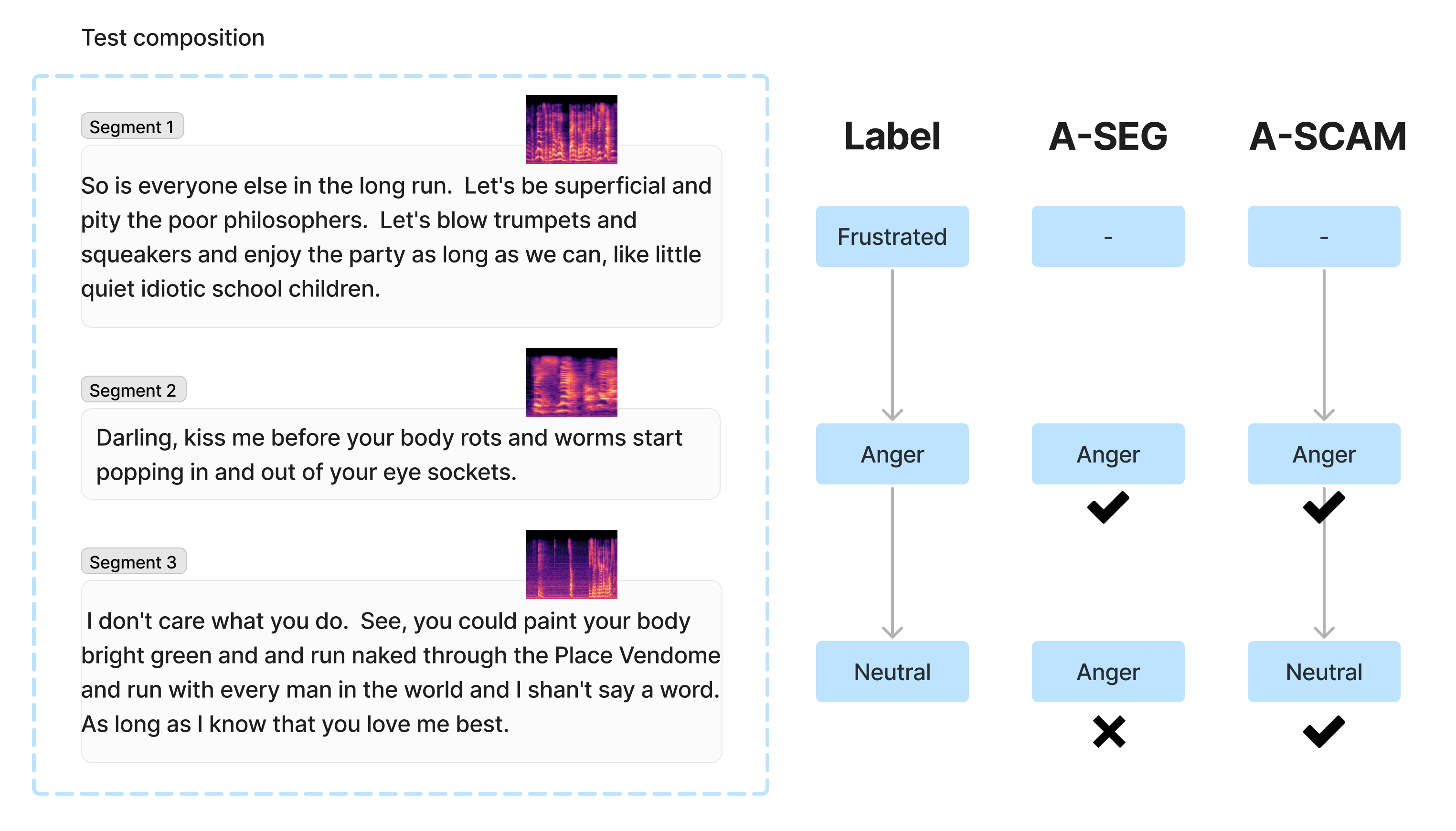}
    \caption{Sample of test composition} 
    \label{Test composition}
\end{figure}

It can be observed that, with the application of SCAM, the performance of the auditory modality improves by 9.36\%, the visual modality improves by 3.79\%, and the multimodal approach improves by 1.45\%. This further demonstrates the effectiveness of SCAM across different modalities. Furthermore, in Figure \ref{Test composition}, visualization of a composition highlights SCAM's remarkable contextual awareness, independent of the consistency in contextual labels. Even when the context undergoes continuous changes, SCAM correctly identifies emotions.

\subsubsection{Multimodal}

\begin{figure}[htbp]
    \centering
    \includegraphics[width=0.4\textwidth]{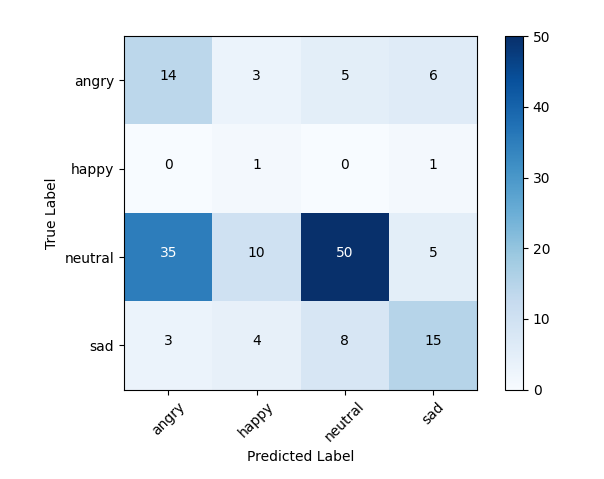}
    \caption{A-SCAM correct, but V-SCAM wrong} 
    \label{adv_speech}
\end{figure}

\begin{figure}[htbp]
    \centering
    \includegraphics[width=0.4\textwidth]{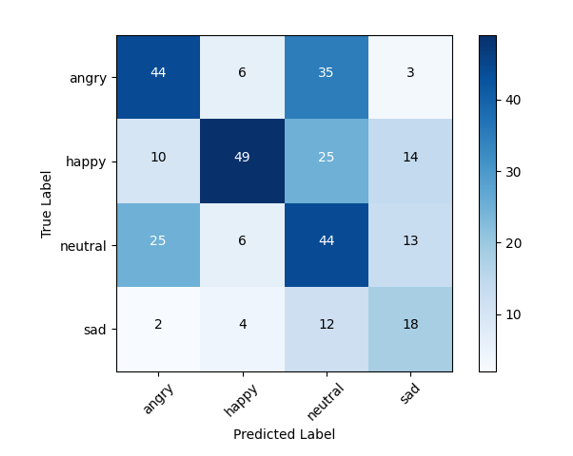}
    \caption{V-SCAM wrong, but A-SCAM correct} 
    \label{adv_visual}
\end{figure}

In general, multimodal performance is expected to surpass unimodal performance. However, in our experiments, the multimodal performance is not superior to the unimodal performance. Therefore, we conduct further analysis of the classification results for the auditory and visual modalities. In the confusion matrix in Figure \ref{adv_speech}, the elements on the diagonal represent correct classification by A-SCAM but incorrect classification by V-SCAM. The remaining entries indicate how V-SCAM incorrectly classifies emotions into different categories. Similarly, in Figure \ref{adv_visual}, we represent cases where V-SCAM classifies emotions correctly but A-SCAM classifies them incorrectly. 

It can be observed that the visual modality is better at recognizing happy, while neutral and angry have a considerable number of correctly classified samples independently by both modalities. This suggests that the two modalities obtain different features for neutral and angry emotions. In some cases, the auditory modality incorrectly identifies angry as neutral and neutral as angry, but the visual modality correctly identifies them, and vice versa. Furthermore, both modalities' primary errors are concentrated in misclassifying some other emotions as neutral, and samples misclassified as neutral differ significantly. Based on the error distribution, the errors made by both modalities are quite similar. 

\section{Conclusion}

In this work, we introduce SCAM, a method that leverages the user's emotion context and features during long-term human-robot interactions for emotion perception. Additionally, we innovatively combine continuous emotion models with discrete emotion models, anchoring the relationships between different emotions using valence and arousal, achieving outstanding performance. Through ablation experiments, we further demonstrate that SCAM significantly improves accuracy in emotion recognition, valence regression, and arousal regression in auditory, visual, and multimodal modalities. Moreover, through data visualization, SCAM performs effectively even in scenarios with continuous changes in context. 

In future work, we will further collect data on robots to validate the reliability of the methods and conduct psychological experiments to analyze the usability of robot emotion perception. Regarding the multimodal conflicts arising from the similarity in probability distributions between the auditory and visual modalities, we will consider introducing additional modal information for emotion perception.

\end{document}